\documentclass{aa}

\usepackage{txfonts}
\usepackage{graphicx}
\usepackage{natbib}
\usepackage{mathrsfs}
\bibpunct{(}{)}{;}{a}{}{,}

\begin{document}

\title{Star formation in the vicinity of the IC 348 cluster}

\author{M. Tafalla \inst{1}
\and M. S. N. Kumar \inst{2}
\and R. Bachiller \inst{1}
}

\institute{Observatorio Astron\'omico Nacional (IGN), Alfonso XII 3, E-28014 Madrid,
Spain
\and
Centro de Astrof\'{\i}sica da Universidade do Porto, Rua das Estrelas, 
4150-762 Porto, Portugal
}

\offprints{M. Tafalla \email{m.tafalla@oan.es}}
\date{Received -- / Accepted -- }

\abstract
{}
{We present molecular line observations of the southwestern part of
the IC 348 young cluster, and we use them together with NIR and
mm continuum data to determine the distribution
of dense gas, search for molecular outflows, and analyze the ongoing 
star formation activity in the region.}
{Our molecular line data consists of
C$^{18}$O(1--0) and N$_2$H$^+$(1--0) maps obtained 
with the FCRAO telescope at a resolution of about $50''$
and CO(2--1) data obtained with the IRAM 30m telescope at
a resolution of $11''$.
}
{The dense gas southwest of IC 348 is concentrated in two groups 
of dense cores, each of them with a few solar masses of material
and indications of CO depletion at high density.
One of the core groups is actively forming stars, while the other
seems starless. There is evidence for at least three
bipolar molecular outflows in the region, two of them powered by previously 
identified Class 0 sources while the other is powered by 
a still not well characterized low-luminosity object.
The ongoing star formation activity is producing a small stellar 
subgroup in the cluster. Using the observed core characteristics
and the star formation rate in the cluster, we propose that
that similar episodes of stellar
birth may have produced the subclustering seen in
the halo of IC 348.
}
{}


\maketitle

%

\section{Introduction}

IC 348 is one of the most studied young clusters. Its stellar population 
has been observed at different wavelengths from the IR to the X rays
\citep{lad95,luh98,her98,naj00,mue03,pre04},
and its surrounding gas, part of the Perseus molecular cloud,
has been mapped with different resolutions in mm-wave lines 
and continuum \citep{bac86,hat05,eno06,rid06,sun06,kir06}. The cluster lies 
at a distance of 320 pc (e.g., \citealt{her98}) 
and consists of more than 300 stars distributed 
with a core-halo structure over a region
20 arcminutes in diameter \citep{mue03}. Superposed 
to the smooth distribution of stars, there is a population
of stellar groups (``subclusters'') each containing 5-20 stars within 
a radius of 0.1-0.2 pc \citep{lad95}. Most stars in IC 348
have formed at a close-to-constant rate 
over the last 2-3 Myr, although some cluster members may be significantly 
older \citep{her98,luh98}.

Star formation in IC 348 seems to have ceased toward the center
but it continues at some level near its southern border,
where the cluster meets the molecular cloud. \citet{str74} identified
an IR object to the SW of IC 348 (IC348-IRS1 hereafter),
and further observations in the visible and IR by \citet{bou95} have
resolved it into a bipolar nebula likely due to an embedded star with 
a disk. \citet{bac87} identified several dense cores 
in its vicinity, and \citet{mcc94} found additional 
signposts of recent star formation in the form of shock-excited 
H$_2$ emission. The most prominent of these H$_2$ features is the 
HH211 flow, which originates from source HH211-MM and is 
associated with a highly collimated molecular outflow mapped
in CO and SiO \citep{gue99,cha01,pal06,hir06}. Another prominent H$_2$
feature has been recently associated with an outflow
by \citet{eis03}, who found an additional mm source 
that seems to power it. The NIR
and optical observations of \citet{eis03} and \citet{wal05,wal06} 
reveal multiple HH objects in the region and suggest that additional
young stellar objects (YSOs) lie embedded in the dense gas.

In this paper we present the results from a survey of the
southwest region of IC 348 (IC348-SW hereafter) aimed
to study its star formation activity and the relation between 
the dense gas and the molecular outflows. These data reveal the
presence of several aggregates of dense cores, some
of them starless, together with at least three molecular outflows
(two newly mapped in CO), and help clarify the kinematics 
of both the outflow gas and the dense material.
Combining our observations with published IR data, we show that 
the subclusters in the halo of IC 348 first identified by \citet{lad95} 
could have originated from star formation episodes 
inside small core aggregates 
like the one currently active in IC348-SW.

\section{Observations}

\begin{figure*}
\centering
\includegraphics[width=11cm, angle=270]{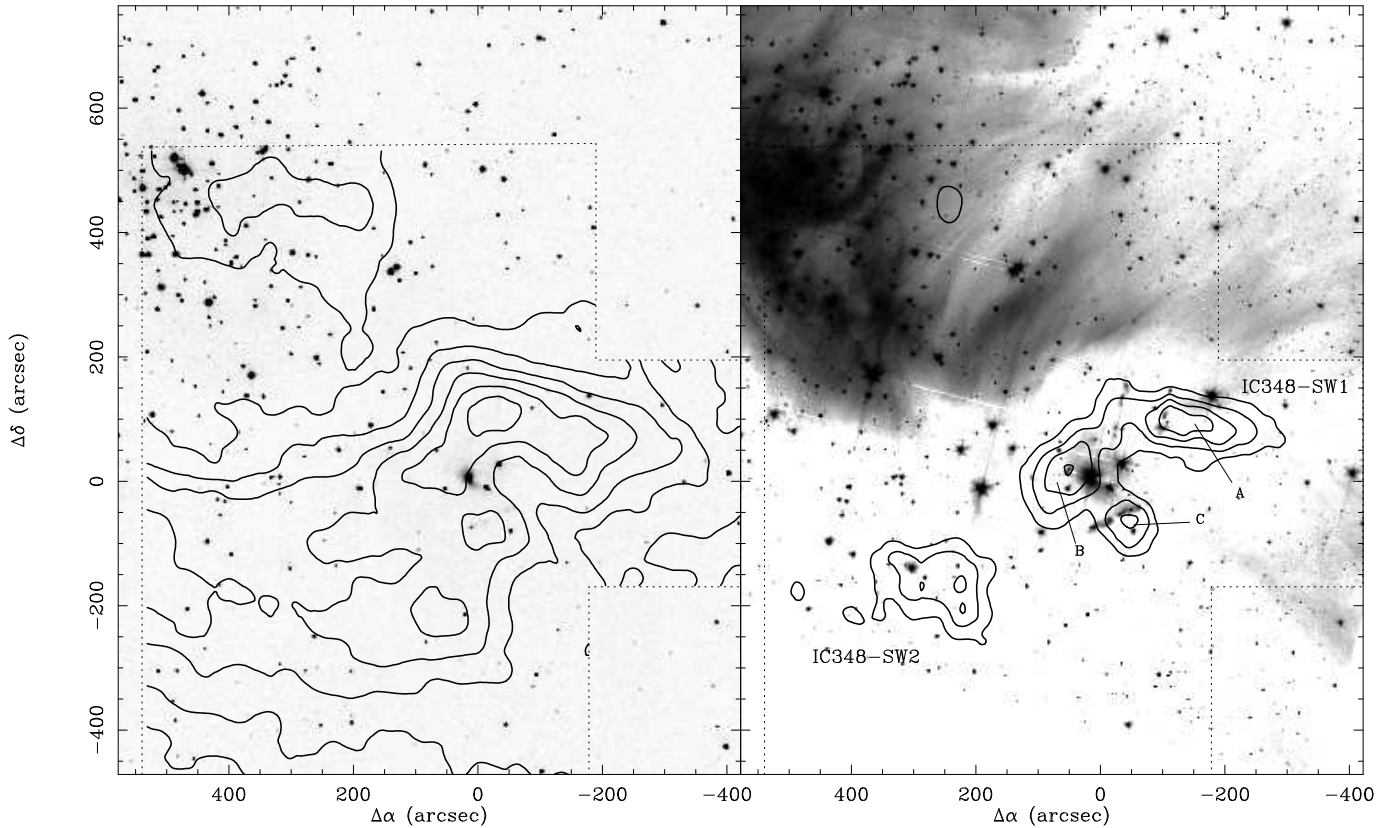}
\caption{FCRAO maps of the molecular gas in the south-western part of the
IC 348 cluster. {\em Left: } C$^{18}$O(1--0) integrated 
intensity map in contours superposed on the K-band 2MASS image. The 
C$^{18}$O emission forms an arc that partly surrounds the IC 348 cluster
(located at the top left). {\rm Right: } N$_2$H$^+$(1--0) emission 
(contours) superposed on the Spitzer IRAC1 (3.6 $\mu$m) image. 
The dense gas traced by N$_2$H$^+$ forms two groups of dense cores,
IC348-SW1 and IC348-SW2, and a small condensation toward the
cluster center. The map center is at $\alpha_{2000}$=03:43:58.8,
$\delta_{2000}$=32:01:51, and the
first contour and interval is 0.5 K km s$^{-1}$ for C$^{18}$O(1--0)
and 1 K km s$^{-1}$ for N$_2$H$^+$(1--0). The dashed line indicates the 
region covered by the FCRAO observations. The C$^{18}$O(1--0) map has
been convolved to a final resolution of $65''$ (FWHM) to filter out
high frequency noise.
}
\end{figure*}

We observed the IC348-SW region in C$^{18}$O(1--0) and
N$_2$H$^+$(1--0) with the (then) 16-pixel SEQUOIA array
receiver on the FCRAO\footnote[1]
{FCRAO is supported in part by the National Science Foundation
under grant AST 94-20159, and is operated with permission of the Metropolitan
District Commission, Commonwealth of Massachusetts.}
telescope in 2001 April. The observations were done in
frequency switching mode using a correlator that provided
velocity resolutions of 0.03 km s$^{-1}$ (C$^{18}$O) and
0.06 km s$^{-1}$ (N$_2$H$^+$). The pointing was checked and corrected
using observations of SiO masers, and the data were converted
into the main-beam temperature scale using an efficiency of
0.55. The FWHM of the telescope beam was approximately $50''$.

We observed the vicinity of IC348-IRS1 in the 1.2mm continuum with the
MAMBO1 bolometer array at the IRAM 30m telescope in 1999 December.
One on-the-fly map was made with a scanning speed of $4''$ s$^{-1}$,
a wobbler period of 0.1 s, and a throw of $41''$. The atmospheric
optical depth was estimated from sky dips carried out immediately
before and after the map, and the data were reduced with the NIC 
software
package. We also observed the IC348-SW region simultaneously in 
CO(1--0) and CO(2--1) with the IRAM 30m telescope in 2000 October and 2001
October. The observations were done in dual-polarization, position
switching mode and were centered at $\alpha_{2000}$=03:43:58.8,
$\delta_{2000}$=32:01:51. The reference position was 20 arcminutes
north from the map center, and it was checked to be free from
detectable emission in the velocity range of interest.
A correlator split into four sections 
provided velocity resolutions of 0.20 and 0.41 km s$^{-1}$ for
the 1--0 and 2--1 spectra, respectively. The atmosphere
was calibrated by observing ambient and cold loads, and
standard efficiency values were used to convert the telescope intensities
into the main-beam brightness scale. The pointing was checked 
making cross scans on bright continuum sources, and
the FWHM of the telescope beam was $21''$ 
at the CO(1--0) frequency and $11''$ at the CO(2--1) and 1.2mm
continuum frequencies.

Near-infrared observations were made using the 3.8m UKIRT with the UIST
array camera on the night of 2002 December 5 under excellent seeing
conditions ($0.5''$ in K band). The camera provided a plate scale of
$0.12''$ per pixel with a total field of view of $120'' \times 120.''$ 
We obtained a 9 point jitter of 60 second exposure time in the K band. 
Standard procedures
for data acquisition and reduction were followed, involving flat
fielding, sky, and dark subtraction of the raw frames. The limiting
magnitude of the image was 17.6.

\section{Large-scale distribution and dense cores}

Figure 1 shows in contours our large-scale FCRAO maps of the C$^{18}$O(1--0)
(left) and N$_2$H$^+$(1--0) (right) emission toward the south-west
vicinity of the IC 348 cluster superposed on the 2MASS K-band 
and Spitzer IRAC1 (3.6 $\mu$m) images (see \citealt{jor06} for
full IRAC maps of Perseus).
As the C$^{18}$O map shows, most of the molecular material lies along 
an arc whose center approximately coincides with the center of the 
IC 348 cluster and has a radius of about $12'$ or 1 pc (see \citealt{hat05} 
for a complete C$^{18}$O(1--0) map of the region). The dense gas, traced 
by N$_2$H$^+$(1--0) in the right panel, consists of two groups of cores
along the C$^{18}$O arc plus a weaker core to the north, close to 
the IC 348 center. The western group of cores, which we will refer to
as IC348-SW1, was mapped previously 
in NH$_3$ by \citet{bac87}. It has an approximate horseshoe shape
and consists of three cores named A, B, and C by \citet{bac87} 
(see Fig. 1). Some of these cores contain well known young stellar
objects (YSOs): core B is associated with IC348-IRS1 
\citep{str74}, core C contains HH211-MM
\citep{mcc94,gue99}, and between cores B and C lies
the mm source IC348 MMS of \citep{eis03}. The second group of
cores is located to the southeast of IC348-SW1 and will be referred to as 
IC348-SW2. It consists of two cores, none of them
associated with a known YSO or an IRAS source. As Fig. 1 shows, both
IC348-SW1 and SW2, and their surrounding molecular
gas, coincide with an  almost total absence 
of scattered light at 3.6 $\mu$m. This suggests that the 
two regions lie in the front part of the IC 348 cluster.

The N$_2$H$^+$ emission of IC348-SW1 in Fig. 1 
is remarkably similar to the NH$_3$ emission mapped by
\citet{bac87} and to the continuum emission
mapped by \citet{hat05} at 850~$\mu$m and \citet{eno06}
at 1.1~mm. Such a good agreement
between maps suggests that these tracers 
reflect the true distribution of
dense gas in the region. The C$^{18}$O(1--0) emission, on the 
other hand, shows little contrast over the map and
and peaks at a different position.
The brightest C$^{18}$O(1--0) is located
between the A and B cores of IC348-SW1,
and the IC348-SW2 group of N$_2$H$^+$ cores coincides with a region of 
relatively weak C$^{18}$O emission. 
This contrast between the C$^{18}$O maps
and the maps of N$_2$H$^+$, NH$_3$, and the continuum 
is unlikely to result from optical depth effects, as
the C$^{18}$O(1--0) lines are less than 4 K in intensity
and therefore not optically thick
(the thick CO(2--1) lines are typically 20 K bright).
It most likely results from C$^{18}$O being strongly depleted at the 
highest densities. C$^{18}$O (and CO) depletion 
due to freeze out onto dust grains is a common phenomenon
in the cold, low-turbulence cores of clouds like Taurus,
were it is by now well characterized
(e.g., \citealt{cas99,taf02}). Its finding 
in the warmer and more turbulent cores of IC348-SW 
indicates that CO depletion occurs at a relatively broad range of 
conditions.

To estimate the mass of each dense core we use the N$_2$H$^+$ emission. 
We first estimate the central H$_2$ column density $N(H_2)$ by
assuming an N$_2$H$^+$ excitation similar to the one found in Taurus,
and an N$_2$H$^+$ abundance of $1.5 \times 10^{-10}$ (also
as found in Taurus, e.g., \citealt{taf04}). 
This method produces H$_2$ column densities for cores A, B, and C
that agree within a factor of 2 with the continuum-based 
estimates by \citet{eno06}, and we take this factor as
indication of the level of uncertainty of our estimate.
We then measure the
core radius $R$ from the N$_2$H$^+$ map and assume an internal density structure
of a critical Bonnor-Ebert sphere (e.g., \citealt{alv01}). In this way, 
the mass of the core is given by 
$$M = 1.6 \; \left( {N(\mathrm{H}_2) \over 10^{22} \; \mathrm{cm}^{-2}} \right)
\left( {R \over 0.1 \; \mathrm{pc}} \right)^2 \; M_\odot.$$
With this approximation, we derive masses of
9, 7, and 3 M$_\odot$ for cores A, B, and C,
respectively, which are 
in reasonable agreement with the NH$_3$-based mass estimates
of \citet{bac87}.
For the IC348-SW2 group of cores, we derive masses of 5 M$_\odot$ (western
core) and 1.5 M$_\odot$ (eastern core), and for the core close to 
the center of the IC 348 cluster, we estimate an approximate mass
of 1 M$_\odot$.

\begin{figure}
\centering
\resizebox{8cm}{!}{\includegraphics{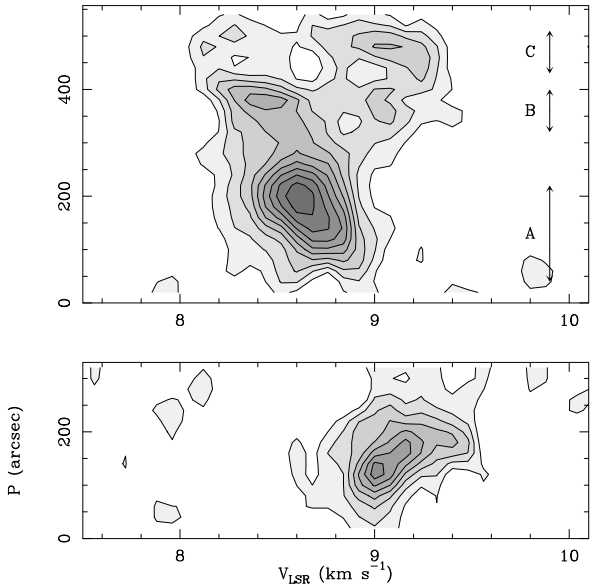}}
\caption{Position-velocity diagrams of the N$_2$H$^+$(JF$_1$F=101-012) 
(``isolated component'')
emission toward IC348-SW1 (top) and IC348-SW2 (bottom).
For IC348-SW1, the position axis follows a curved path that
intersects the peaks of the A, B, and C cores, and the 
the approximate 
location of each core is indicated with arrows to the right. 
For IC348-SW2, the position axis follows a straight line
with PA = 60$^\circ$. In both panels, the first contour and interval 
are 0.1 K.
}
\label{fig2}
\end{figure}

We have explored the velocity structure of IC348-SW1 and SW2 using the 
N$_2$H$^+$(1--0) data, and we illustrate the main features with 
the two position-velocity (PV) diagrams of Fig. 2. The PV diagram for
SW1 follows a curved path along the horseshoe, and the
diagram for SW2 follows the long axis 
of the region (PA = 60$^\circ$). As the top panel shows, the
A core presents a single velocity component with a small velocity
gradient ($\approx 1.3$ km s$^{-1}$ pc$^{-1}$)
and a peak at around 8.6 km s$^{-1}$. The B and C cores,
on the other hand, present lines with two peaks separated by about 
0.5 km s$^{-1}$, being brighter the
blue component in core B and the red component in core C. The presence of
these two components seems not to result from an overlap
between the cores, as each core appears distinct and centrally 
concentrated in
the map of Fig. 1. In addition, the weaker red component of core B 
peaks at the same position as the brighter blue component, as if the 
two were correlated, and a similar but weaker correlation can be seen
in core C. This behavior suggests that in each core the two components
have a common origin. As we will see in the next section, the B and C cores
are affected by outflows that have already
accelerated part of the ambient gas (as seen in CO). It is therefore
likely that the N$_2$H$^+$ components arise
from outflow acceleration, as it has been previously found
in other systems like L1228 \citep{taf97}.

The velocity structure of SW2, on the other hand, is simpler than 
that of
SW1. The N$_2$H$^+$(1--0) line presents a single peak over the region,
and there is a velocity gradient again at the level of 
1.3 km s$^{-1}$ pc$^{-1}$. This velocity gradient 
could represent a smooth change in the central velocity 
over SW2 or result from the two cores of SW2
having different velocities at the level of 0.2 km s$^{-1}$.
The contrast with the double-component velocity structure of
SW1 may result from the SW2 region having no
evidence for star formation and or outflows (see below).

\begin{figure}
\centering
\resizebox{8cm}{!}{\includegraphics{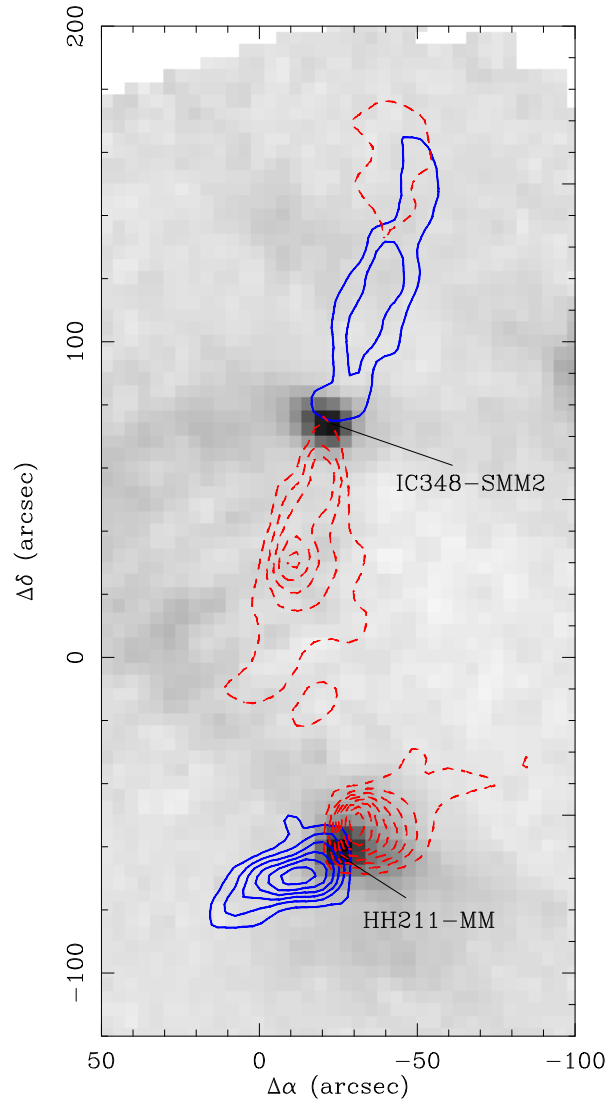}}
\caption{Fast outflows in IC348-SW1. The contours show the CO(2--1)
emission between 5 and 25 km s$^{-1}$ with respect to ambient
cloud (at $V_\mathrm{LSR} = 8.6$ km s$^{-1}$). The solid blue 
contours represent blueshifted
gas and the dashed red contours represent redshifted gas.
The grey scale shows the 1.2 mm continuum emission. The two
bright mm sources at the intersections of the red and blue contours are 
IC348-SMM2 (top) and HH211-MM (bottom). Offsets as in Fig. 1, and 
first contour and contour interval is 3 K km s$^{-1}$.
}
\label{fig3} 
\end{figure}

\section{Molecular outflows}

We have searched for molecular outflows the IC348-SW1 and IC348-SW2
regions using CO(1--0) and CO(2--1) observations. No outflows were 
found towards IC348-SW2, while
at least three outflows (two newly detected in CO)
were found in IC348-SW1. To simplify
the presentation, we separate the outflow velocity regime
into ``fast'' and ``slow.''

\subsection{Fast gas}

We define as ``fast'' the gas that moves at more than 5 km s$^{-1}$
with respect to the ambient cloud (ambient $V_\mathrm{LSR} = 8.6$ km s$^{-1}$ 
in IC348-SW1). This 5 km s$^{-1}$ value corresponds to
ten times the typical FWHM of the N$_2$H$^+$
line in IC348-SW1, and therefore guarantees the absence of
contamination by ambient emission. The exact choice of the
value, however, has little effect on the following discussion.

As Fig. 3 shows, the fast regime
is dominated by two bipolar outflows, each centered on
a bright 1.2mm compact source. The southern outflow is the
well-known HH211 system, first identified by \citet{mcc94} 
from its H$_2$ emission, and later mapped at high resolution
in CO and SiO by \citet{gue99}, \citet{cha01}, \citet{pal06}, 
and \citet{hir06}. 
The mm peak at its center have been variously referred as 
HH211-mm by \citep{mcc94} or SMM1 \citep{wal06}.
The second fast outflow coincides with the extended H$_2$ emission 
detected by \citet{mcc94} (their H$_2$ ``chain'') and \citet{eis03}
(their ``region 1''), and it also coincides with the HH797 A and B 
features recently observed in [\ion{S}{ii}] by \citet{wal05}.
Compared with the optical and IR data, our CO observations
provide velocity information, and show that the outflow
is bipolar with respect to the northern mm source (referred as 
IC348 MMS by \citealt{eis03} and as SMM2 by \citealt{wal06}).
The CO data, therefore, shows unambiguously that SMM2
powers the outflow (we will follow the \citealt{wal06}
notation in the following discussion).

\begin{figure}
\centering
\resizebox{8cm}{!}{\includegraphics{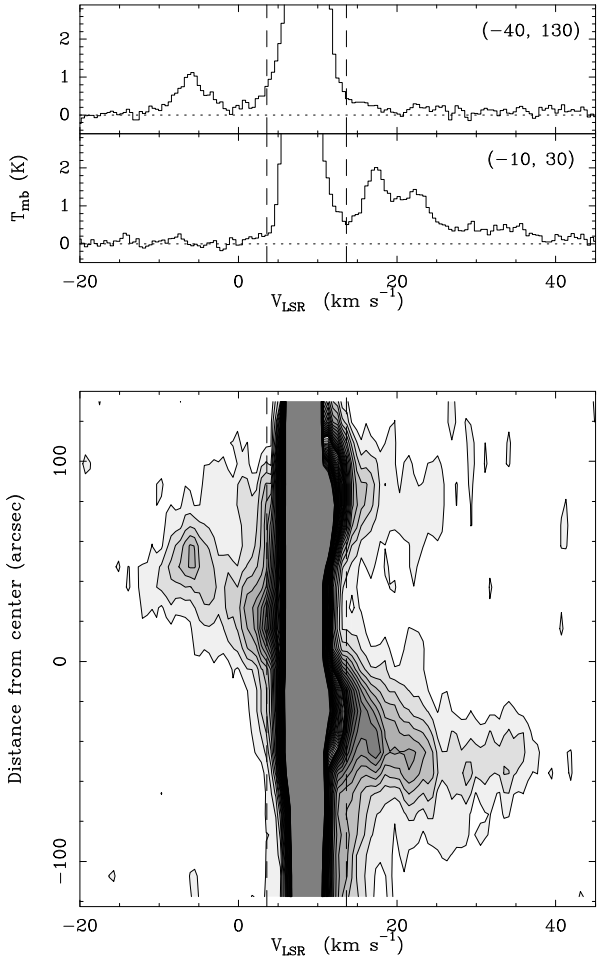}}
\caption{CO(2--1) spectra and position velocity (PV) diagram along
the IC348-SMM2 outflow. The spectra present secondary
peaks of high velocity emission (``bullets'') at positions
symmetrically located with respect to SMM2.
The PV diagram shows how the outflow speed
increases with distance from SMM2 and how the bullets
are highly localized in both velocity and position.
Note that the offset in the spectra are
referred to the map center (Fig. 1), so SMM2 is at ($-25''$, $75''$),
while in the PV diagram, the position axis 
measures distance from SMM2.
The dashed lines mark the boundary of fast regime. First contour
an interval in the PV diagram are 0.1 K.
}
\label{fig4}
\end{figure}

The HH211 and SMM2 outflows present several differences and similarities.
HH211 is more compact than the SMM2 outflow, and although this
may indicate an intrinsic difference between the outflows it could also
result from a projection effect. Indeed, the SMM2 outflow
presents some mixing of blue and red emission at low velocities (see below), 
which suggests that the flow runs close to the plane of the sky.
In any case, both outflows 
seem to belong to the class of highly collimated flows. The interferometric 
data of \citet{gue99} (also \citealt{cha01,pal06,hir06}) show that the fastest 
part of the HH211 outflow is unresolved with arcsecond beams. 
The SMM2 outflow, on the other hand, is unresolved by our $11''$ single-dish
beam at the highest speeds. In addition, its CO spectra present at some 
positions secondary peaks of high velocity (see Fig. 4). These peaks, often 
called ``bullets,'' are
commonly associated with jet-like components in outflows \citep{bac90},
and their presence in the SMM2 system suggest the existence of 
extremely collimated gas. Interferometric observations of the SMM2 
outflow are necessary to further characterize this component.

\subsection{Slow gas}

\begin{figure*}
\centering
\resizebox{15cm}{!}{\includegraphics[angle=270]{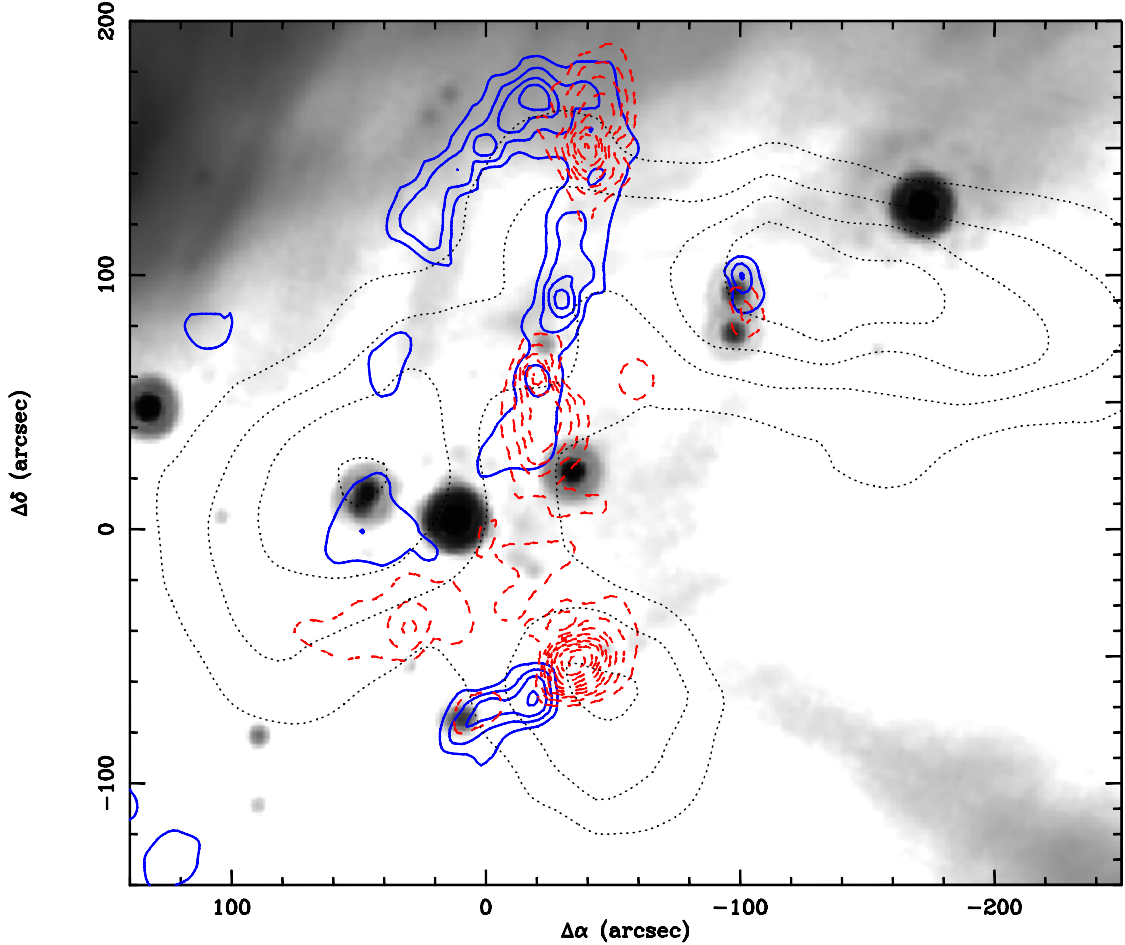}}
\caption{Slow outflow gas toward IC348-SW1. 
The contours show the CO(2--1) emission integrated between
3 and 5 km s$^{-1}$ with respect to the ambient
cloud (at $V_\mathrm{LSR} = 8.6$ km s$^{-1}$).
The solid blue lines represent blueshifted emission, 
and the dashed red lines represent redshifted emission.
The black dotted contours indicate the N$_2$H$^+$(1--0) emission
also shown in Fig. 1, and the grey-scale image in the background
is the 24 $\mu$m Spitzer MIPS image from the c2d Legacy Project.
At least three CO bipolar outflows are present in the region (see
text). Offsets as in Fig. 1, and 
first CO contour and contour interval is 1 K km s$^{-1}$.
}
\label{fig5}
\end{figure*}

The threshold of 5 km s$^{-1}$ used for the fast gas guarantees
the absence of contamination by low-velocity material, but limits our
sensitivity to weak or slow outflows. As a second step in our
outflow search, we now explore the gas moving at velocities lower than
this threshold, but still avoiding contamination from ambient gas.
We have experimented with different velocity ranges and have found
that the gas moving between 3 and 5 km s$^{-1}$ from the
ambient cloud still seems free from ambient contamination. For
this reason, we will refer to this range as the ``slow gas'' regime,
although it should be noted that this gas still
moves supersonically with respect to the ambient cloud.

Figure 5 shows the slow gas regime of CO(2--1) toward the IC348-SW1
region (solid blue and dashed red contours)
superposed to the N$_2$H$^+$(1--0) emission (dotted lines) and the 
Spitzer Space Telescope 24 $\mu$m image from the Cores to Disks (c2d) 
Legacy
Project \citep{eva03}. The HH211 and IC348-SMM2 outflows are still
the dominant elements of the slow CO emission, but a number of new features
are visible. In the IC348-SMM2 outflow, there is a red component at 
the end of the northern blue lobe and a blue arc to its northeast. 
The origin of these components is suggested by their location with 
respect to the dense gas traced by N$_2$H$^+$. 
The N$_2$H$^+$ emission shows that the SMM2 outflow 
runs between the A and B cores 
and that the two anomalous features occur at the edge of 
the region, where 
the outflow leaves the dense gas and encounters the surrounding
medium. The anomalous red emission continues the direction of the flow and 
coincides with a region where the shocked H$_2$ emission becomes brighter
and broader \citep{mcc94,eis03,wal06}. This suggests that the anomalous
red CO emission results from a broadening of the outflow when it 
encounters the surrounding lower
density medium, as it has been observed in other systems 
like the Orion HH212 outflow
\citep{lee00}. The outflow broadening may result from a change in the 
gas regime from isothermal to adiabatic, and if the outflow
moves close to the plane of the sky, it can naturally produce
the observed superposition of blue and red gas (e.g., \citealt{cab88}).
The blue arc to the east, on the other hand, has a less clear origin.
It could result from a strong deflection of the outflow, but
there is no evidence for an obstacle along the outflow path. In fact, the 
wind responsible for the outflow seems to continue
unimpeded past the region with anomalous blue emission. This is
inferred from the long and highly collimated chain of H$_2$
emission, which continues
at least up to the bright ``1-w'' knot of \citet{eis03}. Such a knot
is located at offset ($-54''$, $243''$) with respect to our map center,
and therefore is well aligned with the collimated CO outflow 
past the region of anomalous blue gas (it is in fact so much further 
north from the blue gas that it lies outside our map). A more likely
interpretation, therefore, is that the blue gas is not related to the
outflow. Fig. 5 shows that the blue gas forms an arc parallel to the 
dense gas distribution traced by N$_2$H$^+$ and that approximately points 
towards the center
of the IC 348 cluster (about 1 pc away in projection). We have seen
from the C$^{18}$O emission (Fig. 1) that the cluster seems to have 
excavated a circular hole in the molecular gas, so it is possible
that the blue CO emission results from this destructive action of the
cluster and not from the CO flow. This would also explain 
naturally the shift to the blue, as the IC348-SW1 region is located
slightly in front of the cluster (Section 3), so the dense gas will
be pushed from behind.

After the CO emission associated directly with the HH211 and
SMM2 outflows, the next brightest peak of the slow regime
occurs near $(30'', -40'')$ and is part of a 
region of red gas that lies along the southwestern edge of core 
B (Fig. 5). This red gas is not continuously connected to the
red lobe of SMM2, but the presence of another region of red 
emission near $(-20'', -20'')$ and a number of nearby H$_2$ knots
in the images of \citet{wal06} suggests that the two pieces
of red gas are indeed connected and associated with the southern lobe of 
the SMM2 outflow. If so, the southern lobe would be
approximately as long as the northern lobe.

About $40''$ north of the bright red peak just discussed, there is 
a triangular patch of blue gas whose northern vertex coincides
with the ``region 3'' of H$_2$ emission in the study of \citet{eis03}.
As Fig. 5 shows, this region also presents bright 24 $\mu$m emission,
which suggests the presence of an additional embedded object.

\begin{figure}
\centering
\resizebox{8cm}{!}{\includegraphics{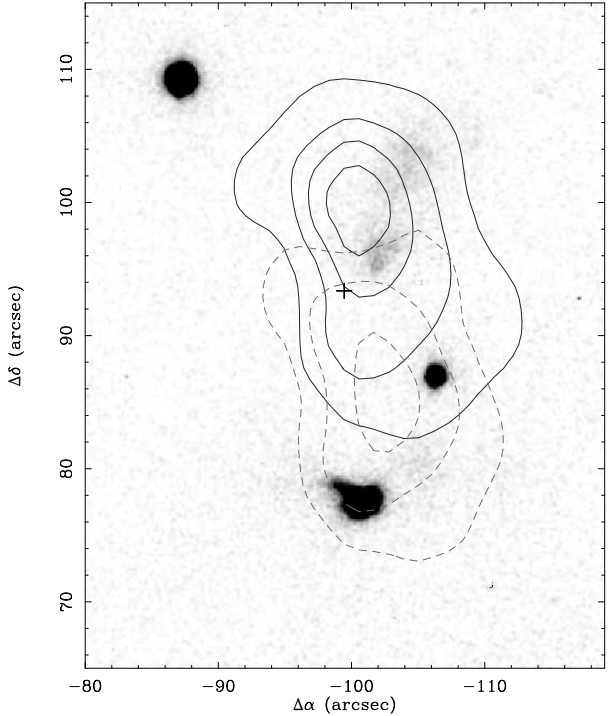}}
\caption{IC348-SMM3 outflow. The solid (blue) and dashed (red) 
contours indicate the CO(2--1) emission blueshifted and redshifted
between 3 and 5 km s$^{-1}$ with respect to the ambient cloud (slow
outflow regime). The grey scale shows our UKIRT K-band image
and the plus sign indicates the position of a 24 $\mu$m source.
Offsets as in Fig. 1, and first contour and interval is
1.0 K km s$^{-1}$. 
}
\label{fig6}
\end{figure}

A better defined feature of the slow
CO emission occurs near $(-100'', 100'')$. There, a blue
and a red lobe of gas form a very compact outflow
at the eastern edge of core A. This newly found CO flow
runs approximately north-south, parallel to the SMM2 outflow,
and seems associated with the ``region 2'' of H$_2$ emission
identified by \citet{eis03}. The close-up view 
in Fig. 6 shows that the outflow has diffuse
H$_2$ emission both towards its north and south lobes, and that
its center lies about $8''$ to the northeast of 2MASS source
03435056+3203180. The 2MASS source, however, is probably not
the outflow source, as its JHK colors are consistent with those 
of a background or TTauri star \citep{lad92}. A more likely
candidate for the exciting source is provided by the $24\mu$m Spitzer 
image. As Fig. 5 shows, there are two sources superposed to the 
CO emission,
one toward the geometric center of the outflow (plus sign in Fig. 6)
and the other coinciding with the bright H$_2$ knot at the end of
the red lobe. Both 24 $\mu$m sources are different from the 2MASS source
and are classified as YSO candidates in the current
c2d catalog \citep{eva05}. The northern source,
close to the outflow center, is most likely the 
driving source of the outflow, while the second source may result 
not from an embedded object but from line emission in the shock.
A similar situation seems to occur toward the HH211 outflow, where 
the end of the blue lobe coincides with
a bright 24 $\mu$m peak (Fig. 5). This 24 $\mu$m peak is also
classified as a YSO candidate in the c2d catalogue, but 
it has no evidence for mm or submm emission (e.g., Fig. 3), 
while its position
coincides with the point of strongest IR and optical shock
emission in HH211 (e.g., \citealt{wal06}). Line emission 
from outflow shocks near 24 $\mu$m is in fact expected from theoretical
grounds. \citet{hol89} predict that the 
[\ion{S}{i}] line at 25.2 $\mu$m and the [\ion{Fe}{ii}] line at 
26.0 $\mu$m will be among the brightest 
atomic fine-structure lines in shocks at densities
between $10^4$ and $10^6$ cm$^{-3}$, typical of dense
cores (their Fig. 7), and these two lines fall inside
the 4.7 $\mu$m-wide bandpass of the 24 $\mu$m MIPS detector. 
Bright [\ion{S}{i}] emission at 25.2 $\mu$m
has in fact been observed by \citet{nor04} toward the
Cepheus E outflow, which is also powered by a very young
object, so this emission is also expected from the outflows in the
IC348-SW1 region. Until spectroscopic observations of the
24 $\mu$m sources are available, their association with 
a YSO remains uncertain 
(the lower wavelength Spitzer bands of IRAC can also
be contaminated by H$_2$ emission, \citealt{jor06}).
A more secure counterpart for the exciting source of the compact 
outflow is provided
by the submm continuum emission observed by different authors.
Source 15 of \citet{hat05}, source 102 of \citet{eno06},
and source SMM3 of \citet{wal06} all coincide with the
approximate center of the CO outflow, and even our
1.2 mm observations, where the source is close to the map edge,
show evidence of emission at the 100 mJy level. Given this
more secure identification of the source at submm wavelengths,
from now on we will refer to the outflow driving source as IC348-SMM3.

\subsection{Outflow energetics and nature of the powering sources}

\begin{table}
\caption[]{Outflow energetics
\label{tbl-1}}
\[
\begin{array}{lccc}
\hline
\noalign{\smallskip}
\mbox{Outflow~~~~~~~~~~~~~} & \mbox{Mass~~~~~} & \mbox{Momentum~~~}  & \mbox{Energy~~~}
\\
& \mbox{(M$_\odot$)} & \mbox{(M$_\odot$ km s$^{-1}$)} & \mbox{(erg)} \\
\noalign{\smallskip}
\hline
\noalign{\smallskip}
\mbox{HH211-MM} & 1.5\;10^{-2} & 1.5\;10^{-1} & 6\;10^{43} \\
\mbox{IC348-SMM2} & 3.3\;10^{-2} & 3.1\;10^{-1} & 1\;10^{44} \\
\mbox{IC348-SMM3} & 10^{-3} & 4\;10^{-3} &  4\;10^{41} \\
\noalign{\smallskip}
\hline
\end{array}
\]
\end{table}

We calculate the mass, momentum, and energy of the outflows 
using the CO(2--1) emission, which we assume is optically thin in the
outflow
regime and that has an excitation temperature of 20 K, as suggested by the
ambient peak intensities (these two assumptions make our estimate a lower
limit). We also assume a CO abundance of $8.5 \; 10^{-5}$ \citep{fre82}
and make no correction for outflow inclination. Neglecting the contribution
of gas moving at less than 3 km s$^{-1}$ from the ambient cloud
($V_{\mathrm{LSR}}=8.6$ km s$^{-1}$), we obtain the outflow parameters
listed in Table 1. 
Although the HH211 outflow has been studied before, previous estimates
of its parameters were based on interferometer data,
and therefore missed a significant fraction of the emission. The 
Table 1 values, therefore, contain the first full estimate of
the HH211 energetics.

As can be seen in the table, the HH211 and SMM2 outflows have very
similar parameters. Their mass, momentum, and energy
differ by a factor of 2 or less, which is little
given all the uncertainties involved. The slightly 
larger values of the
SMM2 flow (despite its location closer to the plane of the sky) may
arise from the larger luminosity of the source: 8 L$_\odot$ versus
the 4 L$_\odot$ of HH211-MM \citep{fro03, eis03}.
These similar energetics, together
with the similar speeds and high collimation, suggest that the
powering sources of the HH211 and SMM2 outflows
share similar physical properties and evolutionary state. 
\citet{fro03} and  \citet{eis03} have classified these sources
as Class 0 \citep{and93}, and the outflow parameters in Table 1
confirm that the two objects belong to 
the youngest protostellar phase.

The IC348-SMM3 outflow, on the other hand, has a very small 
energy content, close to the detection limit of our survey. Compared
with the HH211 and SMM2 outflows, the SMM3 outflow is about 
one order of magnitude less massive and has two orders of
magnitude less kinetic energy. These numbers, together with the
small spatial extent of the CO emission (about $30''$ or
0.04 pc) suggest that the flow is powered by a source of much lower
luminosity than HH211-MM or SMM2. The exact luminosity of SMM3, 
unfortunately,
is not well constrained. The 24 $\mu$m flux may be contaminated
by outflow emission (see above), and no data close to the
peak of the SED are available yet (the IRAS data are
confused by the extended emission from the IC 348 cluster). 
Although the outflow
parameters suggest that SMM3 is a low luminosity source, it is 
unlikely that its energy output is as low as that of
the very low luminosity object (VELLO) L1014-IRS 
(0.09 L$_\odot$, \citealt{you04}). For this object, \citet{bou05} 
estimate an outflow mass 
about 70 times smaller and a kinetic energy at least two orders of
magnitude lower than those of the SMM3 outflow. Although the 
interferometric data used to derive the L1014-IRS outflow
parameters only provide  a lower limit because of filtering
of the extended emission, the difference in values is too
large to be explained as an artifact. It seems therefore likely 
that SMM3 is more luminous than 0.09 L$_\odot$.

\section{Star formation in IC348-SW and its relation to 
the IC 348 cluster}

We have seen that the IC348-SW1 group of cores 
is relatively rich in young stellar objects.
Sources HH211-MM, IC348-SMM2,
and IC348-SMM3 all drive bipolar outflows and are therefore YSOs 
likely formed in IC348-SW1 within the last $\approx 10^5$ yr 
(Class 0 objects are even younger, \citealt{and93}).
The IC348-IRS1 object,
whose discovery provided the first indication that star formation was
still ongoing in IC 348 \citep{str74}, has a 
less clear status. It has been interpreted as
a late B star with an edge-on disk 
by \citet{bou95}, but recent IR observations by \citet{luh98}
suggest that it has an early M spectral type. Its
large extinction \citep{luh98} and 
reflection nebulosity \citet{bou95} indicate youth, but the lack of
a well defined outflow (Section 4.2) suggests that it is more evolved
than the other three sources; a complete spectral energy distribution
is still needed to properly classify this YSO. Other young
objects in the region have been identified in the surveys of
\citet{her98} and \citet{luh98}, who searched for evidence of 
H$_\alpha$ emission and IR excess. Among these objects, IfAHA 7 and 8 
stand out for lying projected onto the N$_2$H$^+$ emission 
and having H$_\alpha$ equivalent widths of classical T Tauri 
stars (35 and 30 \AA\ for IfAHA 7 and 8, respectively, \citealt{her98}). Given 
their expected age ($\approx 10^6$ yr, \citealt{luh98}), however, 
it is unclear whether these stars have originated from the IC348-SW1 
group of cores or they belong to a more distributed stellar
population that has already drifted from its natal place. Even
discounting these older objects, the presence of 4 YSOs,
two of them of Class 0, makes IC348-SW1 the most active star-forming 
site of the IC 348 cluster. 

The study of IC348-SW1 may help elucidate the process of star formation
in the IC 348 cluster. \citet{lad95} have found that the stellar distribution
in the outer cluster presents significant substructure. In addition to 
a central core of stars with a radius of 0.5 pc, the authors identify eight
subclusters, each containing 5-20 sources within a radius of 
0.1-0.2 pc. The sizes of these subclusters match the approximate size of the
IC348-SW1 and SW2 regions as measured from the N$_2$H$^+$
maps (radii 0.1-0.3 pc), and this suggests a connection between 
the two types of objects. Using the 4 YSOs identified
in IC348-SW1, which lie inside a circle of $85''$ radius, we 
crudely estimate a stellar density of 70 stars pc$^{-2}$. 
This density is already half the mean stellar density of the Lada \& 
Lada subclusters, and equals the density of their subcluster ``c.''
Adding the nearby sources IfAHA 7 and 8 \citep{her98} and sources
49, 160, and 124 from the \citet{luh99} catalog further enriches the
stellar census toward IC348-SW1, to which new sources will be 
incorporated if the cores form additional
stars with the dense gas still available. The final result of star
formation in IC348-SW1 and its vicinity will therefore be a stellar
density enhancement similar to the subclusters found by
\citet{lad95}. 

To explore whether star formation in regions like
IC348-SW1 is a viable mechanism to produce the observed subclustering,
we need to investigate whether the number of 
stars found in subclusters is consistent with their
production rate and dispersal
time scale.  A given group of stars will appear as a distinct 
density enhancement until its members drift apart due to
proper motions, and for the case of IC348-SW1 we can estimate
this timescale from the typical velocity difference between its 
constituent cores (the linewidths
may be contaminated by core disruption, see section 3). From our
N$_2$H$^+$(1--0) spectra, we estimate that the typical core-to-core
velocity difference is less than
0.3 km s$^{-1}$, in agreement with the NH$_3$ estimate
of \citet{bac87}. This line-of-sight estimate implies a plane of 
the sky dispersion close to 0.4 km s$^{-1}$ if the velocity field 
is isotropic. As the subclusters have a
typical radius of 0.15 pc \citet{lad95}, stars will travel a diameter
distance in about 0.75 Myr, which we take as a typical
stellar group lifetime. To calculate the number of stars formed
in this time, we assume that the cluster has been forming stars
at a constant rate over the last 3 Myr \citep{mue03},
and that half of its 380 stars have formed in the outer cluster \citep{lad95}.
Thus, over the last 0.75 Myr, about 47 stars have formed in 
the outer IC 348 cluster. This number should be compared
with the 82 stars found by \citet{lad95} in the subclusters of the
halo, but first we need to subtract the
contamination by the more diffuse cluster population. Dividing the
number of non subcluster objects in the outer cluster by the area of this
region, we estimate that the contamination fraction is 1/3 (density 
of diffuse population is 50 stars pc$^{-2}$ and mean density in a subcluster
is 150 stars pc$^{-2}$). This implies that 
there are about 55 stars in the subclusters of the outer IC 348,
which is roughly consistent with the 47 stars expected if the stars
formed in regions like IC348-SW1. The subclustering structure, therefore, 
could have resulted from star formation
in aggregates of dense cores like the ones we have studied. Indeed, if
star formation in these loose aggregates produces a different
proportion of low mass stars than the formation in a more compact
environment, likely responsible for the central stellar core of IC 348, 
the difference may help explain the radial dependence of the IMF
found by \citet{mue03}. 

Finally, we note that the velocity dispersion between aggregates of 
cores seems to be larger than the velocity dispersion between 
the cores of a given aggregate.  Our N$_2$H$^+$
mapping of the IC 348 vicinity is not complete enough to determine 
the full statistics of the aggregate-to-aggregate kinematics, but it already
provides some interesting hints. The SW1, SW2 aggregates and the
core towards the IC 348 center have mean
LSR velocities of 8.6, 9.1, and 7.9 km s$^{-1}$, respectively, from
which we estimate a (necessarily crude) rms value of 0.6 km s$^{-1}$
together with a trend for the velocity difference to increase with distance
between aggregates. This dispersion is larger than the difference
of about 0.3 km s$^{-1}$ found between the cores of IC348-SW1, and
suggests that most of the initial velocity dispersion between the stars in
the core will result from the relative velocities of the aggregates
and not from the small scale turbulence inside the aggregates. It
also suggests that the stars formed from a given aggregate may
keep a distinct velocity pattern for about 0.75 Myr, which
should be testable with a proper motion study of the subclusters.

\section{Summary}

We have mapped the molecular gas south-west of the IC 348 young
cluster in C$^{18}$O(1--0), N$_2$H$^+$(1--0), and CO(2--1).
Combining these data with observations of the young stars
and YSOs in the NIR and mm continuum, we have reached the following
conclusions.

1. The dense gas south-west of IC 348 is concentrated in two
regions, referred here as IC348-SW1 and IC348-SW2, each containing
several cores of a few solar masses of material. The two regions
are embedded is lower density gas and they seem located in the front
part of the IC 348 cluster. While IC348-SW1 is actively forming
stars, IC348-SW2 seems quiescent.

2. The distribution of N$_2$H$^+$(1--0) emission is similar to the
distribution of NH$_3$ and submm continuum mapped by previous 
authors, but differs significantly from that of C$^{18}$O(1--0).
As the rare isotope is not optically thick, the disagreement
suggests that the CO molecule is depleted at the core centers
(likely due to freeze out onto grains) as previously seen in 
the more quiescent cores of Taurus.

3. The fast CO emission ($|\Delta V| > 5$ km s$^{-1}$ with respect
to the ambient cloud) is dominated by two highly collimated outflows,
each associated with a bright mm/submm source. The first outflow
is the molecular counterpart of HH211 and has already been mapped
using interferometers by a number of authors. The second outflow,
not previously mapped in CO, is associated with a previously
known chain of H$_2$ emission and several HH objects, and is referred
here as the IC348-SMM2 outflow. The characteristics and energetic
content of the two outflows are consistent with them being excited
by Class 0 sources.

4. At low velocities ($3 < |\Delta V| < 5$ km s$^{-1}$),
the CO emission presents additional features associated
with the SMM2 outflow and (probably) the interaction of the 
cluster and the dense gas. In addition, we find a compact
outflow associated with with a mm/submm object (SMM3) also
detected at 24 $\mu$m. The small size and low energy
content of this new outflow suggests that it is powered
by a source of rather low luminosity.

5. The ongoing star formation activity in IC348-SW1 
is producing a significant stellar density enhancement,
similar to others previously identified in the halo of the IC 348
cluster from NIR observations. From an estimate of the production
rate and dispersal timescale of aggregates like  
IC348-SW1, we conclude that similar episodes of star
formation could have originated
the substructure seen in the IC 348 cluster.

\begin{acknowledgements}

We thank Chris Davis for obtaining the infrared data of
SMM3 through a service program at UKIRT, Jennifer Hatchell
for providing us with her 850$\mu$m data of the region,
and Jorge Grave for creating a color version of Fig. 1.
MT and RB acknowledge partial support from grant AYA2003-7584.
This research has made use of NASA's Astrophysics Data System
Bibliographic Services, the SIMBAD database, operated at CDS,
Strasbourg, France, and the Two Micron All 
Sky Survey, which is a joint project of the University of Massachusetts 
and the Infrared Processing and Analysis Center/California Institute of 
Technology, funded by the National Aeronautics and Space Administration 
and the National Science Foundation. The DSS
was produced at the Space Telescope Science Institute under US Government
grant NAG W-2166. This work is based in part on observations made with 
the Spitzer Space Telescope, which is operated by the Jet Propulsion 
Laboratory, California Institute of Technology under a contract with NASA.

\end{acknowledgements}

\end{document}